\documentclass[referee,fleqn,usenatbib]{mnras}

\usepackage{graphicx}	

\title[The impact of shocks in Nitrogen abundances]{The case of NGC 6302: The impact of shocks in the derivation of Nitrogen abundances.}

\author[Lago P .J .A et al.]{
Lago, P.J.A.,$^{1,2}$\thanks{E-mail: plago@lna.br}
Costa, R.D.D.,$^{2}$
Fa\'undez-Abans M.,$^{1}$
Maciel, W.J.$^{2}$
\\
$^{1}$LNA-Laborat\'orio Nacional de Astrof\'isica, R. dos Estados Unidos, 154 - Na\c{c}\~oes, Itajub\'a-MG, Brazil\\
$^{2}$Department of Astronomy, IAG, Universidade de S\~ao Paulo, Rua do Mat\~ao, 1226 - Cidade Universit\'aria S\~ao Paulo-SP, Brazil\\
}

\date{Accepted XXX. Received YYY; in original form ZZZ}

\pubyear{2018}

\begin{document}
\label{firstpage}
\pagerange{\pageref{firstpage}--\pageref{lastpage}}
\maketitle

\begin{abstract}

High nitrogen abundance is characteristic of Type I planetary nebulae as well as their highly filamentary structure. In the present work we test the hypothesis of shocks as a relevant excitation mechanism for a Type-I nebula, NGC 6302, using recently released diagnostic diagrams to distinguish shocks from photoexcitation. The construction of diagrams depends on emission line ratios and kinematical information. NGC 6302 shows the relevance of shocks in peripheral regions and the importance to the whole nebula. Using shocks, we question the  usual assumption of ICF calculation, justifying a warning to broadly used abundance derivation methods. From a kinematical analysis, we derive a new distance for NGC 6302 of $805\pm143\,$ pc.
\end{abstract}

\begin{keywords}
radiation mechanisms: non-thermal -- ISM: abundances -- ISM: kinematics and dynamics
\end{keywords}



\section{Introduction}

The study of gas kinematics in planetary nebulae (PNe) is very important since it provides valuable information about their formation and evolution through time. The kinematics shapes their structure, dictating the morphology. The origin of PNe is adequately described by the Generalized Interacting Stellar Winds (GISW) theory \citep{1987AJ.....94..671B}, an upgraded version of the Interacting Stellar Winds (ISW) theory \citep{1978ApJ...219L.125K},
that emphasizes the importance of interactions between distinct gas ejection episodes. Besides the ionization of the circumstellar gas by the strong UV radiation of the central star (CS), shock interactions also provide a mechanism that brings energy into the gas, being highly related with kinematics, as it is a consequence of gas interaction and the dissipation of its mechanical energy.

Shocks have been used to explain the  morphological components and spectroscopic characteristics for many astrophysical objects, such as Herbig-Haro objects \citep{1998AJ....116.2943R}, supernova remnants \citep{2019MNRAS.482.5268Z}, cataclysmic variables \citep{2014MNRAS.442..713M}, symbiotic stars \citep{2009MNRAS.396..807C}, and were recently related to molecular emission in extragalactic sources \citep{2019arXiv190104273G}. In the context of PNe, shocks have also been considered as a possible excitation mechanism \citep[e.g.][]{2009ApJ...700.1067M, 2016MNRAS.455..930A} to explain some of their intriguing morphological features such as the low-ionization structures (LISs); \citep{2001ApJ...547..302G}. In order to provide an alternative interpretation for observed features, in section 4 we will present a proper discussion on this topic.

The derivation of abundances seems to be unrelated with shocks, but assuming that the shocks could change the emission line fluxes and the derived abundances, it is natural to suppose that shocks should be considered. It is impossible to derive the chemical abundances directly from observational data since for most the species the fluxes from all ions are not usually available.  A possible way to derive such abundances is by the use of empirical factors, the Ionization Correction Factors (ICFs), presented for the first time by \citet{1969BOTT....5....3P}. In the literature, there are several ICF sets and suggestions based on simulations \citep{2014MNRAS.440..536D}. Here we present for the first time an estimate of the impact of shocks on the nebular abundances.

The kinematics of gas in PNe and similar nebulae, as well as its geometry  have been studied by several authors using high-resolution spectroscopy and/or proper motion measurements \citep[e.g.][]{1984MNRAS.210..341S,2011MNRAS.416..715S,2012ApJ...761..172G,2014ApJ...797..100F,2014Msngr.158...26S,2015A&A...582A..60C}. In order to model the objects in these studies, the combination of spectroscopy and photometry is crucial, in view of the great advantage of minimizing the projection effects.
Considering this background we present an analysis of shocks as excitation/ionization mechanisms, combining data from our own observations, virtual observatories and literature. The goal is to investigate the role of shocks in excitation and ionization of the planetary nebula NGC 6302. The interpretations applied to some microstructures are used to the nebula as a whole. New information is derived from the data, such as the distance to NGC 6302. A possible relationship between kinematics and alterations in the abundance calculation is also explored in the ICF context.

In section 2 the data used in the study are presented, showing specific features of each source and instrumentation, in section 3 the results are described, section 4 discusses the main results and in section 5 a summary of the work is given.

\section{The Data}

\subsection{Observations and data reduction}

\begin{figure*}
\begin{center}
\includegraphics[scale=0.64]{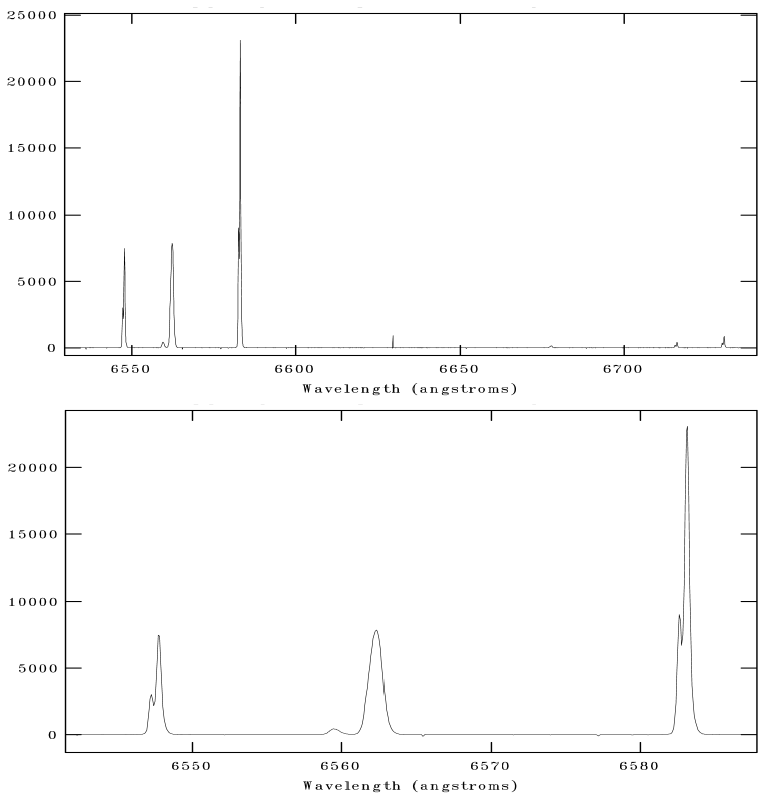}
\caption{Example of OPD spectra, it is possible to see the [NII] and [SII] emission lines. The bottom panel is a zoom, the double peak in the forbidden lines is due to the gas kinematics. The separation of these components were derived from a Gaussian fit, that allowed the derivation of expansion law and values are presented in the appendix. The spectra were not flux calibrated.}
\label{fig:coude}
\end{center}
\end{figure*}

New data for NGC 6302 were collected at Pico dos Dias Observatory (OPD) MCTIC/LNA, in Brazil. The Perkin-Elmer 1.6m telescope was used with a Coud\'e spectrograph that provides a spectral resolution of $11\,$km/s using a $1800\,$l/mm grating at first direct order of diffraction. Observations of NGC 6302 were performed in May 2012.  Slit position was kept along the symmetry axis of the nebula. Each exposure required 20 to 30 minutes. The seeing during the observations varied between 1.5 and 2 arcsecs. An example of the extracted spectra is shown in Fig ~\ref{fig:coude}.

The data were bias subtracted and flat-field corrected, wavelength-calibrated using a Th-Ar lamp, and corrected by systemic velocity by adopting $-35.7\,$kms$^{-1}$ \citep{1953GCRV..C......0W}. Data reduction was made using IRAF package (Image Reduction and Analysis Facility)\footnote[2]{http://iraf.noao.edu/}; cosmic rays were removed manually to avoid contamination, also using IRAF.

In order to study the variation of kinematic properties, several apertures were used along the slit (positioned on the symmetry axis of the nebula). These apertures covered all nebula extension. The size of the aperture was kept larger than the seeing, the chosen size was $2"$, and the position presented in the appendix table is related to the apertures and they were selected using the Apall task from IRAF. The aim is to acquire the kinematic information for each aperture, in order to use it to compare distinct regions of the nebula and its excitation conditions. The separation of the blueshift and redshift components, for each aperture, allowed the construction of the appendix table, as well as the distance derivation.

\subsection{Virtual Observatories (VO's)  and literature data}

For this work, various data from the literature have been used.
The emission line maps of  \citet{2014A&A...563A..42R} for NGC~6302 have been used to get an estimation of several emission line ratios involving the lines [N II], [O II], [S II] and H$\alpha$ as well as the values of the electronic density. Using SOAR telescope, \citet{2014A&A...563A..42R} built 11"x250" maps for NGC 6302, with angular resolution of $1.45$"x1".
These authors have observed and computed several emission lines for maps (a sample of twenty five), summarized in table 1 of their paper. The electronic density was estimated by the ratios [S II] 6716/6731 and [Ar IV]  4711/4740, as well as the electronic temperature was estimated by [O III] (4959 + 5007)/4363 and [N II] ( 6548 + 6584)/5755  ratios\footnote{The wavelengths are given in Angstroms in these ratios}.

Images from the Hubble Space Telescope (HST) database\footnote{Availeble on https:https://hea-www.cfa.harvard.edu/USVOA/science-tools-services/, The US Virtual Observatory Alliance - All HST images were obtained by this portal} were used as a reference for morphological features, since they have the best angular resolution available. Nevertheless, we did not use the HST images to derive the line ratio maps due to the bandwidth of the filter F656N. This filter, that covers H$\alpha$ mainly, is too wide, so the contamination from the [NII] line (654.8nm) cannot be neglected.

\section{Results}

\begin{figure*}
\begin{center}
\includegraphics[scale=0.42]{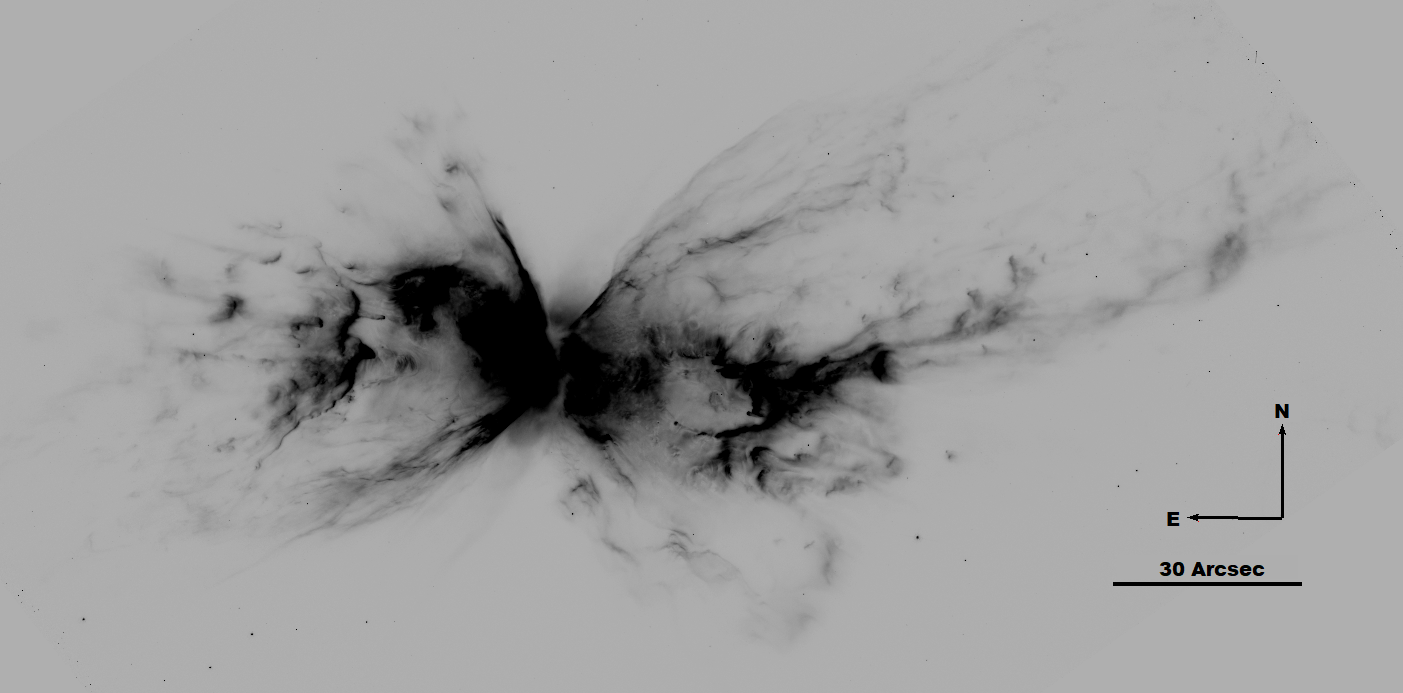}
\caption{Image from HST, filter 658N. Note the structure of NGC 6302, where substructures can be clearly seen.}
\label{fig:image6302}
\end{center}
\end{figure*}

NGC 6302 is classified as a Type-I PN, following the classification scheme of \citet{1983IAUS..103..233P}, being part of the classical study.  The most important characteristics are the pronounced filamentary structure, as can be seen in Figure ~\ref{fig:image6302}, and the high nitrogen abundances. The bipolarity of NGC 6302 is clear and this image is taken as our reference for a sequence of assumptions in the paper.

\citet{2011MNRAS.416..715S} made a full kinematical map of one lobe of NGC 6302 , but to build this map they used the proper motions of the gas, giving a velocity law in mas/year. Since a distance-independent velocity law was desirable, our own data from OPD, described in section two, were used to derive the velocity law by assuming that both lobes are identical and the velocity field follows a linear law. We also adopted  an inclination relative to the plane of the sky of $12.8^{\circ}\pm2^{\circ}$ for the nebula from \citet{2008MNRAS.385..269M}.

Our assumptions are based on Figure \ref{fig:image6302} and literature data. Even considering that in small scale each lobe displays a unique filamentary structure, their large scale structures can be considered as similar. Concerning the hypothesis of linearity in the velocity field, it is well known that this nebula displays a linear velocity field, even if the simulations are still not able to reproduce it \citep{2014MNRAS.442.3162U}.

In order to  derive the expansion law for the NGC 6302 lobes, a Gaussian profile was adjusted for each emission line, deblending the components of each line. Each aperture was corrected to systemic velocity. The expansion law is given as a function \textit{V(r)}, a bipolar geometry was used and the two lobes were considered symmetrical. We have included in the appendix a table with the derived velocities. The typical error for the velocities was 10 km\space s$^{-1}$.

The inclination of the PN was used to derive the velocity in the plane of the sky $V$, since redshifted and blueshifted components are different for each lobe. The observed velocities $V_{obsI}, V_{obsII}$ are the blueshifted and redshifted components for each lobe, respectively, and $\theta$ is the inclination. Therefore we have 

\begin{equation}
V_{obsI}-V_{obsII}=2Vsin(\theta)
\end{equation}

Calculating the value of this difference for each aperture, and using  linear regression, an expansion law could be derived as

\begin{equation}
\rm
V(r)=(1.7\pm0.3 km s^{-1}arcsec^{-1})r+(14\pm5.5km s^{-1})
\end{equation}

where r is the distance to the nebula center in arcsec.

\section{Discussion}

LIS (Low Ionization Structures) are characterized, in a general way, as following the kinematic properties of the surrounding gas, rim shells or halos, except for fast low-ionization emission regions (FLIERs), that are the fast counterparts of LIS, as discussed by \citet{1993ApJ...411..778B,1994ApJ...424..800B}. These structures can be found in all morphological types of PNe, and there is no important temperature contrasts between the LIS and the rest of the PN \citep{2016MNRAS.455..930A}. 

LIS can be divided into four morphological types: knots, filaments, pairs of jets and jet-like features. The morphological diversity of LIS can be associated to the expected origins of such structures, some of them challenging the  proposed theoretical scenario. For a complete characterization of LIS see \citet{2001ApJ...547..302G}, where the proposed models are confronted with observational results. 

Using high angular resolution, \citet{1996A&A...313..913C} have studied LIS in a wide range of PNe of different morphological types. Shocks were not invoked to justify the observed features of LIS, but the image ratio $(H\alpha + [NII])/[OIII]$ was used to identify LIS. In more recent works the ratio  $[OIII]/H\alpha$ was used to characterise shock-defined regions \citep{2007apn4.confE..14R,2008A&A...489.1141R,2013A&A...557A.121G}. The main difference between these line ratios is $(H\alpha + [NII])/[OIII]$ is  insensitive to density variations and focus in ionization conditions variations, $[OIII]/H\alpha$ is sensitive to variations of excitation or ionization. An important consideration should be pointed out: some of cited works are constructed from simulations, and do not consider the reddening, that must impact the line ratios in observational studies.

Nitrogen overabundances related to LIS seem to be irrealistic. Derivation of the nitrogen abundance in PNe generally makes use of an ionization correction factor (ICF), that is $ N^{+}/N=O^{+}/O$. The models of \citet{2006MNRAS.365.1039G} on NGC 7009 predict values of 0.6, 0.72 and 0.62 for $ (N^{+}/N)/(O^{+}/O)$, respectively for the rim, knots and the whole nebula, this difference is due to ionization effects, instead of the consideration of factor 1 from $ N^{+}/N=O^{+}/O$. \citet{2006IAUS..234..405G} also address this point and inquire if the use of such ICF could compromise the classification of the PNe.

Within the context of LIS, \citet{2007apn4.confE..14R} and \citet{2008A&A...489.1141R} performed a series of numerical simulations. In a set of 11 axisymmetric simulations, they simulate a denser bullet moving through an uniform medium, away from the source of an ionizing radiation field. The models are different in the features of the bullet, such as velocity and density, and different source temperatures are considered as well. Results were compared to surface brightness distributions of bow shocks, and several observed properties could be reproduced including shocks as an ionization/excitation mechanism.

The role of shocks in the excitation/ionization context will depend on some environmental features. Higher velocity bullets produce stronger shocks, as well as a denser medium. However, a minimal contrast between the excitation of the radiation field and the shock should exist to allow its detection. \citet{2009ApJ...700.1067M} obtained the same result studying H$_{2}$ emission in cometary knots in the infrared.

\citet{2016MNRAS.455..930A} reinforce that overabundances in LIS are not real, and point to excitation mechanisms to justify the presence of intense low-ionization emission lines. They suggest new diagnostic diagrams to characterise not only LIS but also shock-ionized regions. These diagnostic diagrams are based on other parameters than line ratios, and seem to be efficient to distinguish those regions with relevance of shocks from regions dominated by photoionization.  They are the ratios $[NII]/H\alpha$, $[OII]/[OIII]$, $[OII](372.7\,$nm$)/H\alpha$ and $[SII](672.5\,$nm$)/H\alpha$ against the ratio $f_{shock}/f_{*}$. The parameter $f_{shock}$, is the flux of ionizing photons produced by shocks, defined from the energy flux  $F_{shock}$ by \citet{1996ApJS..102..161D}, in units of $erg$ cm$^{-2}s^{-1}$. $F_{shock}$, the energy flux, can be converted into $f_{shock}$, the photon flux, by dividing the energy flux by the average energy of a photon, L$/S_{*}$; where $S_{*}$ is the photon rate and L is the stellar luminosity. Therefore the ratio $f_{shock}/f_{*}$ can be defined as:

\begin{equation}
f_{shock}/f_{*}=\rm 9.12 \times 10^{-3} (\frac{Vs}{100kms^{-1}})^{3} \times (n_{e}/cm^{-3}) \times \frac{\pi d^{2}}{L}
\end{equation}

where $f_{*}$ is the photon flux from the star at distance d and Vs is the velocity of the shock, We assume that this velocity is the same that can be derived by equation 2.
\citet{2016MNRAS.455..930A} define three regions in their diagnostic diagrams based on the $f_{shock}/f_{*}$ ratio. Shock-excited regions have $log(f_{shock}/f_{*})>-1$, a transition zone is defined by $-2<log(f_{shock}/f_{*})<-1$, and the photoionized regions are those where $log(f_{shock}/f_{*})<-2$. They have shown that, from their sample, only two knots in NGC 6891 are photoionized, advocating the crucial role of shocks in LIS.

Two considerations should be made about the equation 3: The original physical interpretation seems to be inappropriate. The related ratio was appointed as the photon flux relation \citep{2016MNRAS.455..930A}; this assumption depends on the mean energy of the photons from shock. It is unsuitable to assume that the photons from shocks will have the same mean energy from the blackbody photons, and it is also unsuitable to relate it to the CS. Instead of this original approach, we suggest that equation 3 should be interpreted as the ratio of fluxes; the problem of dependence on the mean energy of photon will be suppressed, and the equation remains the same. It will maintain the diagnostic diagrams changing only the interpretation of the ratio. 

The second consideration is about the general use of equation 3. This equation should be faced as a marker of shocks, being useful in the diagnostic diagrams; outside this context, the equation losses its meaning. The shocks models in PNe are still restricted once that the shocks are very complex phenomena with a hydrodynamical background. An adequate analytical expression seems unlikely.

\begin{figure}
\begin{center}
\includegraphics[scale=0.65]{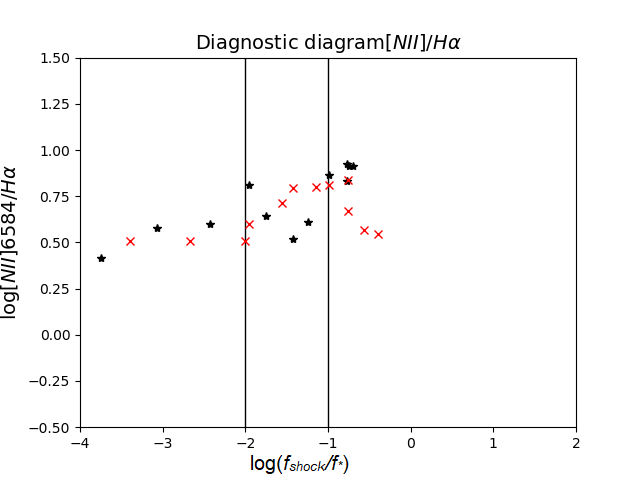}
\includegraphics[scale=0.65]{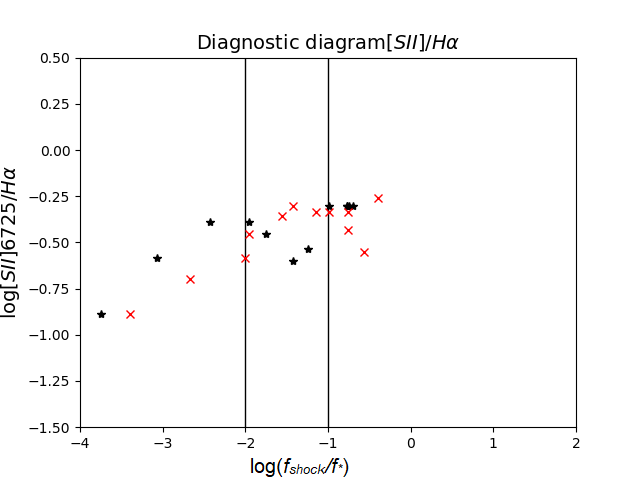}
\includegraphics[scale=0.65]{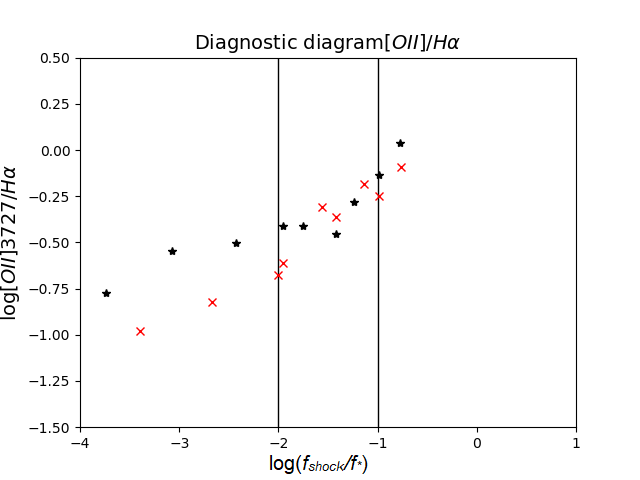}
\caption{Diagnostic diagrams for NGC 6302. Different symbols and colors were used to separate the two lobes of the nebula. The black stars represent the eastern lobe, while the red "x" symbols correspond to the western one. It should be noted that the flux ratios (horizontal axis) have a strong dependence on the distance to the CS:
the expected behavior is to obtain higher ratios with higher distances (see text).}
\label{fig:diag6302}
\end{center}
\end{figure}

Using the observational data and the expansion law, diagnostic diagrams were built for shocks, as can be seen in Figure ~\ref{fig:diag6302} for NGC6302. From figures 3 and 4 of \citet{2014A&A...563A..42R}, we took an offset of 4" from the position of the CS and the physical properties were measured varying $r$ in 4" steps. We chose this spacing in order to optimize the use of data. 

It should be noted that the map of $[OII]/H\alpha$ was not available, so the map of $[OII]/H\beta$ was used instead. For this, the ratio was divided by 2.85, assuming this value for the Balmer decrement. With the adoption of this ratio, reddening effects on the line ratio were also minimized. A luminosity of 5690 L$_{\odot}$ for the CS of NGC 6302 was adopted as a lower limit \citep{2011MNRAS.418..370W}, and the distance used to derive $f_{shock}$ was $1.17\,$kpc \citep{2008MNRAS.385..269M}.

By comparing the expansion law with that derived by \citet{2011MNRAS.416..715S}, it was possible to derive a distance of $805\pm143\,$pc for NGC 6302. Comparing our result with \citet{2008MNRAS.385..269M} it can be seen that we derive the same distance within uncertainties, about 0.14 kpc; the difference is due to the technique used, since we measure it for the whole nebula.

From the diagnostic diagrams for NGC 6302 in Figure ~\ref{fig:diag6302}, one can see that most of the points are in the transition zones or shock zone, supporting the hypothesis of an object with line fluxes strongly affected by shocks. \citet{1991ApJ...367..208L} suggested shocks as a mechanism responsible for high ionization potential emission lines as an alternative scenario instead of a very hot CS, which could not reproduce the characteristics properly. Considering the work of \citet{2016MNRAS.455..930A} and the present results, we argue that shocks seem to be important for the ionization structure studied here. 

The points with smaller $log(f_{shock}/f_{*})$ ratios in Figure ~\ref{fig:diag6302}, which are relative to regions closer to the CS, are found to be dominated by photoexcitation. As mentioned before, the luminosity adopted for the star was a lower limit (5690 L$_{\odot}$), and it is important to keep in mind that the choice of stellar parameters will strongly influence these results; shocks will be hidden by photoionization if the star is considered more luminous. Our choice relies on realistic estimates based on the available observational data.

The line ratios shown in Figure ~\ref{fig:diag6302} also indicate another interesting result. Considering the diagnostic diagrams by \citet{2016MNRAS.455..930A},  it can be seen that the diagram involving the ratio $log[NII]/H_{\alpha}$ is quite different from ours. NGC 6302 shows a considerably higher ratio compared to usual PNe. This higher value is characteristic of Type-I nebulae, and is observed in the whole object. 

It should be noted that, despite the growing use of photoionization models to derive chemical abundances of PN, empirical abundances obtained with the help of ICFs are still very useful for large samples when frequently no information on the CS is available, so  that a clear assessment of the role of shocks in this scenario is important. 

The use of diagnostic diagrams (as used in this work) to a whole well-sampled nebula is a powerful tool to describe the excitation/ionization mechanism. Considering that there is no abundance gradient in the nebula we could investigate the behaviour of ICFs in the two excitation regimes. This is the key to properly investigate the impact of the shock mechanism in PNe, since it is impossible to separate the effects of abundance and different ionization/excitation mechanisms in a heterogeneous sample of nebulae.

Using the diagrams of Figure ~\ref{fig:diag6302}, we can estimate the element abundance relations and check the simulated results by \citet{2006MNRAS.365.1039G}. Considering that the approximation of $N^{+}/N=O^{+}/O$ is correct for photoionization regions \citep{1994MNRAS.271..257K} (KB 94), we can compare this value using our diagrams, by comparing the maximum and minimum values of the line ratios. In a  first approximation it is possible to write

\begin{equation}
log(\frac{[NII]_{shock}}{H\alpha_{shock}})-log(\frac{[NII]}{H\alpha})=0.5
\end{equation}

This equation is derived directly from the diagrams, by subtracting the values in the axis. We are comparing the line ratio in two different regimes, then:

\begin{equation}
\frac{N^{+}_{shock}}{N}\approx 3.16\frac{N^{+}}{N}
\end{equation}

Doing the same for oxygen we find:

\begin{equation}
\frac{O^{+}_{shock}}{O}\approx 5\frac{O^{+}}{O}
\end{equation}

This analysis implies to the following result:

\begin{equation}
\frac{N^{+}_{shock}}{N}\approx 0.6\frac{O^{+}_{shock}}{O}
\end{equation}

A detailed derivation of these equations is given in the Appendix. This result is in agreement with \citet{2006MNRAS.365.1039G}; and points out to inadequate ICFs for not  considering shock mechanisms. To investigate the shock-excited regions, the use of proper ICF is required, otherwise the abundance of nitrogen will be overestimated as has been suggested for LIS \citep{1993ApJ...411..778B}.

Some features of this derivation should be emphasized. The proportionality between ion line fluxes and ionic abundances is considered, even for shocked gas. The classical approach to derive the abundances (by ICFs) uses the same assumptions, but due to different physical characteristics, the method fails, as summarized by equation 7. The result highlights the inadequation of the use KB 94 ICF. When the typical ICF is used, it overestimates the abundance since the method disregards proper physical phenomena. 

Equations 6 and 7 accentuate the effects of reinforcements in line intensities. A factor of 3.16 and 5 multiplies the ratios for Nitrogen and Oxygen, respectively, in peripheral zones. The consequence in ICF for Nitrogen is demonstrated, and the effects on ICFs of other elements will be considered in a future work. Proposing a proper ICF is a tricky task, the relation to the electronic temperature and line intensities is quite complicated, and needs adequate simulations and description of shocks. Equations 7 and 8 next demonstrate that neglecting shocks would result in misjudged values.

The previous result affects new ICFs by \citet{2014MNRAS.440..536D} (DI 14) too. The math remains the same, but the relations between Oxygen and Nitrogen have to change. The new relation will be $\frac{N^{+}}{N}=\frac{O^{+}}{\xi O}$, the term $\xi$ carries two variables implicitly from DI 14. The result will be different from KB 94 by the term $\xi$, depending on the choice of parameters, and the equivalent equation is\footnote{Please check the appendix for more details.}:

\begin{equation}
\frac{N^{+}_{shock}}{N}\approx \frac{0.6}{\xi} \frac{O^{+}_{shock}}{O} 
\end{equation}

\section{conclusions}

Our results show a clear difference between the gas emission in distinct regimes. Due the nature of the Type-I PN considered here, the role of shocks is important in their line fluxes. We have shown that shock mechanism should be studied in a broad perspective to assess the real errors that may affect abundance determinations.

The traditional ICF's recipes consider only  photoionization, but shocks can increase emission lines fluxes, mainly the lines related to lower excitation energies. The results point to an alteration in abundances due to this fact. Therefore,  abundances should not be derived for gas in these conditions. In order to obtain an adequate derivation of the abundances a previous study of the gas in necessary, in order to guarantee that shocks are not modifying the line intensities.

NGC 6302 is a PN whose abundances must be strongly influenced by shocks. Our results indicate a combined effect: the nebula has a large abundance of nitrogen, but shocks are reinforcing these values. 

From a kinematical analysis a distance for this nebula was derived from our data: \mbox{$805\pm143$ pc}.

A detailed study about shocks in a large sample of PNe would be needed in order to clarify their impact in general abundances. The behaviour of this mechanism in Type-II PNe could be used to see whether this impact is restricted to substructures and evaluate the abundances. Another important feature to be considered is how the role of shocks is taken into account in photoionization codes and in the abundances derived from them.

As a final remark photoionization codes, such as MOCASSIN \citep{2003MNRAS.340.1136E} and CLOUDY \citep{1998PASP..110..761F}, are computational tools from which abundances can be estimated through modelling. However, there are no consistent studies relating kinematics and its consequences in abundances. Exists the code MAPPINGS that can simulate photoionization and shocks, such interesting results are presented for PNe (e. g. for Hen 2-111 \citep{2018MNRAS.475..424D}), some tests are done for fast structures in a different approach from the codes previous cited. Until now it is a fact that only photoionization has been considered in the derivation of abundances.

\section*{Acknowledgements}

P.J.A. Lago thanks CNPq for his post-doctoral fellowship (PCI Program - Process 300106/2019-0), and also for his graduate fellowship (Process 140803/2014-9). WJM thanks CNPq (Process 302556/2015-0). We thank FAPESP for the financial support (Process 2010/18835-3). We thank Dr. Denise Gon\c{c}alves and Dr. Stavros Akras for many fruitful discussions, We also thank LNA/OPD for the telescope time granted and support, as well as for USVOA for the availability of the HST data. Finally, We would like to thank the anonymous referee for all comments and suggestions.




\bibliographystyle{mnras}
\bibliography{bibliografia}


\appendix
\section{Math details}

The aim of this section is to improve the comprehension of our results by presenting some details of the math. The first important thing is to face the ICF conditions as dimensionless ratios, so that we can change the fluxes ratios to elements ratios. The only important assumption to keep in mind is that we are assuming proportionality between the quantity of the element and the line flux (relative for the ionized species). 
Let's start with equation 4, repeating it:

\begin{equation}
log(\frac{[NII]_{shock}}{H\alpha_{shock}})-log(\frac{[NII]}{H\alpha})=0.5
\end{equation}

Due to properties of logarithms, it can be rewritten:

\begin{equation}
\frac{[NII]_{shock}}{H\alpha_{shock}}\frac{H\alpha}{[NII]}=3.16
\end{equation}

Then calling the ratio $\frac{H\alpha_{shock}}{H\alpha}=\beta$, we have:

\begin{equation}
\frac{N^{+}_{shock}}{N}\frac{N}{N^{+}}\frac{H\alpha}{H\alpha_{shock}}=3.16
\end{equation}

\begin{equation}
\frac{N^{+}_{shock}}{N}=3.16 \beta \frac{N^{+}}{N}
\end{equation}

Repeating the math for Oxygen and considering proper values we obtain:

\begin{equation}
\frac{O^{+}_{shock}}{O}=5\beta \frac{O^{+}}{O} 
\end{equation}

Remembering that we are using the ICF condition from KB 94, $\frac{N^{+}}{N}=\frac{O^{+}}{O}$, it is possible to substitute the equation A5 inside A4 achieving the result:

\begin{equation}
\frac{N^{+}_{shock}}{N}\approx 0.6\frac{O^{+}_{shock}}{O}
\end{equation}

That was presented as equation 7. The process to achieve the equation 8 is quite similar, the only difference is to admit $\frac{N^{+}}{N}=\frac{O^{+}}{\xi O}$, since the conditions are a consequence of different ICF recipes. 

Starting with equation A4, that is acquired as previously, we can rewrite this as:

\begin{equation}
\frac{N^{+}_{shock}}{N}=\frac{3.16 \beta}{\xi} \frac{O^{+}}{O}
\end{equation}

and finally we have:

\begin{equation}
\frac{N^{+}_{shock}}{N}\approx \frac{0.6}{\xi} \frac{O^{+}_{shock}}{O} 
\end{equation}

That was presented as equation 8.

\section{Velocity table}

\clearpage

\begin{table}
\caption{The table displays the results of the Gaussian fit for the separation of each kinematical component. These velocities were extracted varying the position: the signal was divided into several apertures, to allow the derivation an expansion law. The values presented for velocity are shown in redshift component and blueshift ones, as marked in the top of the column. }
\begin{center}
\label{tablefilter}
\begin{tabular}{lccc}
\hline
\hline
Distance to the centre	& V$_{red}$	&	V$_{blue}$	\\	
(arcsec)&	(kms$^{-1}$)	&	(kms$^{-1}$)	\\	\hline
0	&3.8	&-20.9 \\
2	&3.3	&-18.4 \\
8	&16.6	&-8.8\\
10	&26.5	&-9.9\\
12	&29.0	&-11.5\\
14	&28.9	&-14.9\\
16	&31.1	&-16.9\\
18	&36.9	&-18.3\\
20	&39.8	&-18.8\\
22	&41.7	&-19.3\\
24	&42.2	&-20.2\\
26	&44.1	&-22.2\\
28	&45.2	&-23.4\\
-2	&4.4	&-23.4 \\
-4	&5.0	&-26.5 \\
-6	&6.2	&-29.0 \\
-8	&7.4	&-29.6 \\
-10	&8.7	&-29.6 \\
-12	&10.0	&-29.5 \\
-14	&11.4	&-30.2 \\
-16	&12.1	&-31.8 \\
-18	&12.1	&-33.6 \\
-20	&12.0	&-36.3 \\
-22	&11.5	&-40.0 \\
-24	&11.1	&-40.5 \\
-26	&3.4	&-41.4 \\
-28	&-2.5	&-40.3 \\
-30	&-4.3	&-35.9 \\
\hline
\hline
\end{tabular}
\end{center}
\end{table}




\bsp	
\label{lastpage}
\end{document}